\documentclass[journal,twocolumn]{IEEEtran}

\usepackage{multicol}
\usepackage{etoolbox}
\makeatletter
\patchcmd{\@makecaption}
  {\scshape}
  {}
  {}
  {}
\makeatletter
\patchcmd{\@makecaption}
  {\\}
  {.\ }
  {}
  {}
\makeatother

\usepackage{amsfonts}
\usepackage{mathrsfs}
\usepackage{mathtools}
\usepackage{amsfonts}
\usepackage{amssymb}
\usepackage{graphicx}
\usepackage{epsfig}
\usepackage{psfrag}
\usepackage{amsmath}
\usepackage{array}
\usepackage{cases}
\usepackage{eufrak}
\usepackage{cite,graphicx,amsmath,amssymb,color}
\usepackage{algorithmic}
\usepackage{algorithm}
\usepackage{subfigure}
\usepackage{bm}
\usepackage{multirow}
\usepackage{threeparttable}
\usepackage{array}
\usepackage{makecell}

\newtheorem{Lem}{Lemma}

\newtheorem{Prob}{Problem}

\makeatletter
\renewcommand\normalsize{%
   \@setfontsize\normalsize\@xpt\@xiipt
   \abovedisplayskip 0.025\p@ \@plus0.05\p@ \@minus0.125\p@
   \abovedisplayshortskip \z@ \@plus0.075\p@
   \belowdisplayshortskip 0.15\p@ \@plus0.075\p@ \@minus0.025\p@
   \belowdisplayskip \abovedisplayskip
   \let\@listi\@listI}
\makeatother

\IEEEoverridecommandlockouts

\begin{document}
\bibliographystyle{IEEEtran}

\title{Optimal Multicast of Tiled 360 VR Video}
\author{Chengjun Guo\thanks{Manuscript received May  17, 2017; revised July 20; accepted August 1, 2018. The work of Y. Cui was supported by NSFC grant 61401272 and grant 61521062.
The work of Z. Liu was supported  by JSPS KAKENHI Grant Number JP16H02817 and JP18K18036.
The associate editor coordinating the review of this paper and approving it for publication was M. Velez. \textit{(Corresponding author: Ying Cui.)}
\newline{\indent C. Guo and Y. Cui are with the Shanghai Institute for Advanced
Communication and Data Science,
Institute of Wireless Communication Technologies, Department of Electronic
Engineering, Shanghai Jiao Tong University, Shanghai 200240, China (e-mail:
guochengjun382@sjtu.edu.cn; cuiying@sjtu.edu.cn).}
\newline{\indent Z. Liu is with Department of Mathematical and Systems Engineering, Shizuoka University, Hamamatsu 432-8561, Japan (e-mail: liu@ieee.org).}
},  Ying Cui,~\IEEEmembership{Member,~IEEE}, and Zhi Liu,~\IEEEmembership{Member,~IEEE}}

\maketitle

\begin{abstract}
In this letter, we \textcolor{black}{study optimal  multicast of tiled 360 virtual reality (VR) video
from one \textcolor{black}{server (base station or access point)}} to multiple users.
\textcolor{black}{We consider random viewing directions and random channel
conditions, and adopt time division multiple access
(TDMA).}
\textcolor{black}{For given video quality, we optimize the transmission time and power allocation  to minimize the average transmission energy. For given transmission energy budget, we optimize the transmission time and power allocation as well as the encoding rate of each tile to maximize the received video quality.} \textcolor{black}{These} two optimization problems are challenging non-convex problems.
We obtain globally optimal closed-form solutions of the two non-convex problems, which reveal important design insights for multicast of tiled 360 VR video.
Finally, numerical results demonstrate the advantage
of the proposed solutions.
\end{abstract}

\section{Introduction}

A  \textit{virtual reality} (VR) video is generated by capturing a scene of interest \textcolor{black}{from} \textcolor{black}{all directions} at the same time using omnidirectional cameras. A user wearing a VR headset  can freely watch the scene of interest in any viewing direction at any time, hence enjoying \textcolor{black}{immersive} viewing experience. VR has vast applications in entertainment, education, medicine, etc. It is predicted that the global market of VR related products will reach 30 billion \textcolor{black}{US dollars} by 2020 \textcolor{black}{\cite{111}}.
Transmitting an entire 360 VR video which is of a much larger size than a traditional video brings a heavy burden to wireless networks.
To improve transmission efficiency and avoid view switch delay, a 360 VR video is divided into smaller rectangular segments of the same size, referred to as tiles\textcolor{black}{. The set of tiles} covering a user's current \textit{field-of-view} (FoV) and the FoVs that may be watched shortly should be transmitted simultaneously.

In \cite{unicast-liu,unicast-one}, the authors consider tiled 360 VR video transmission in single-user wireless networks, and optimize video encoding parameters to maximize  the utility which reflects the  quality of the received 360 VR video.
The optimization problems  are discrete, and are solved by exhaustive search.
In \cite{multicast-nojoint,multicast-olcoding}, the authors consider  360 VR video transmission in multi-user wireless networks, and exploit multicast opportunities for serving concurrent viewers to improve transmission efficiency.
In particular,
\cite{multicast-nojoint} adopts the  tiling  technique, and optimizes the modulation and coding level of each tile to maximize the video quality. It is a discrete optimization problem of a large size, and the proposed heuristic  algorithm may not provide desirable performance and complexity.
\textcolor{black}{Different from \cite{unicast-liu,unicast-one,multicast-nojoint}, \cite{multicast-olcoding} considers online encoding} to encode only the required area to be transmitted. \textcolor{black}{In addition,  \cite{multicast-olcoding}} proposes a dynamic multicast mechanism which is  adaptive to the channel conditions of all users. The online encoding in \cite{multicast-olcoding} has higher coding efficiency (at the cost of the increase of implementation complexity). \textcolor{black}{However}, the transmission scheme in \cite{multicast-olcoding} is not optimization based, and may not guarantee promising performance.
Note that the solutions in \cite{multicast-nojoint,multicast-olcoding} do not provide many design insights for 360 VR video multicast in wireless networks. It is still not known how the \textcolor{black}{required} FoVs  and  channel conditions  of all users affect optimal resource allocation.

In this letter, we would like to address the above \textcolor{black}{issues}. We \textcolor{black}{consider optimal  multicast of tiled 360 VR video from one \textcolor{black}{server (base station or access point)}} to multiple users. Specifically, we divide the 360 VR video into tiles. For each user, we deliver a set of tiles that cover the user's current FoV and the FoVs that may be watched shortly.  In order to make use of
multicast opportunities and avoid redundant transmissions, we partition the set of tiles to be transmitted to users into disjoint subsets and multicast different subsets of tiles to different groups of users.
\textcolor{black}{We consider random viewing directions and \textcolor{black}{random} channel conditions, and \textcolor{black}{adopt}   time division multiple access (TDMA).}
\textcolor{black}{For given video quality, we optimize the transmission time and power allocation  to minimize the average transmission energy under the transmission time allocation constraints and the transmission rate constraint. For given transmission energy budget, we optimize the transmission time and power allocation as well as the encoding rate of each tile to maximize the received video quality under the  maximum transmission energy constraint.}
These two optimization problems are challenging non-convex problems.
For each non-convex optimization problem, by analyzing optimality properties, we \textcolor{black}{successfully} transform it into an equivalent convex problem and obtain a globally optimal closed-form solution using KKT conditions.
\textcolor{black}{The derived solutions reveal important design insights for multicast of tiled 360 VR video.}
To the best of our knowledge, these important design insights have never been explicitly explored and analytically verified in existing literature.
Finally, numerical results demonstrate the advantage of the proposed optimal solutions.

\section{System Model}

\begin{figure}[t]
\vspace*{-0.1cm}
\begin{center}
 \includegraphics[width=2.8in]{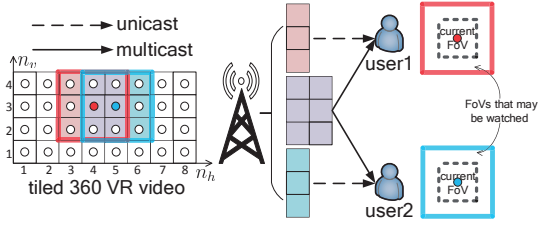}
  \end{center}
  \vspace*{-0.35cm}
     \caption{\small{\textcolor{black}{Illustration} of tiled 360 VR video multicast with $K=2$, $N_h\times\ N_v=8\times4$, $V_h\times V_v=8\times4$. A  circle represents a viewing direction.}}  
\label{fig:system-model}
\vspace*{-0.55cm}
\end{figure}

As illustrated in Fig.~\ref{fig:system-model}, we consider downlink transmission of a 360 VR video from a single-antenna \textcolor{black}{server (base station or access point)} to $K~(\geq1)$ single-antenna users. Let $\mathcal{K}\triangleq\{1,\ldots,K\}$ denote the set of user indices. 
For ease of implementation, \textcolor{black}{as illustrated in Fig.~\ref{fig:system-model},} we consider $N_h\times N_v$  viewing directions, where $N_h$ and $N_v$ represent the numbers of horizontal and vertical viewing directions,
and can be arbitrarily large. 
The $(n_h,n_v)$-th viewing direction refers to the viewing direction in the $n_h$-th row and \textcolor{black}{$n_v$-th} column. When a  VR user is interested in one viewing direction, he can \textcolor{black}{view} a rectangular FoV of size $F_h\times F_v$  with the viewing direction as its center.

To improve transmission efficiency, we consider tiling. In particular, the 360 VR video is divided into $V_h\times V_v$ rectangular segments of the same size, referred to as tiles, where $V_h$ and $V_v$ represent the numbers of segments in each row and each column, respectively.
Each tile is encoded into $L$ versions with $L$ different quality levels and encoding rates.
\textcolor{black}{The $l$-th version of each tile has the $l$-th lowest quality with encoding rate $D_l$. Note that $D_1<D_2<\cdots<D_L$.  For fairness, we assume the tiles transmitted to all users are of the same quality level and let $D$ denote the corresponding encoding rate.
A VR user may watch one FoV at sometime, and freely \textcolor{black}{switch} to another FoV after a while. To avoid view switch delay, for each user, the set of tiles that cover its current FoV and the FoVs that may be watched shortly will be delivered.}

$K$ users \textcolor{black}{randomly} select their viewing directions. Let
$\mathbf{X}_k\in\mathcal X$ denote the random viewing direction of user $k$, where \textcolor{black}{$\mathcal{X}\triangleq\{(n_h,n_v)|n_h=1,\ldots,N_h,n_v=1,\ldots,N_v\}$} represents the set of all possible viewing directions of each user.
Let $\mathbf X\triangleq (\mathbf{X}_k)_{k\in\mathcal K}\in \mathcal X^K$ denote the random system viewing direction state.
For given  $\mathbf{X}$, let $\Phi_k(\mathbf{X})$ denote the set of tiles that need to be transmitted to user $k$, and let
$\Phi(\mathbf{X})\triangleq\cup_{k\in\mathcal{K}}\Phi_k(\mathbf{X})$ denote the set of tiles that need to be transmitted to all users.
In order to make use of multicasting opportunities and avoid redundant transmissions, we divide $\Phi(\mathbf{X})$ into $I(\mathbf{X})$  disjoint non-empty sets $\mathcal{S}_i(\mathbf{X})$, \textcolor{black}{$i\in\mathcal{I}(\mathbf{X})\triangleq\{1,\ldots,I(\mathbf{X})\}$.}
For all $i,j\in I(\mathbf{X}),$ $i\neq j$, $\mathcal{S}_i(\mathbf{X})$ and $\mathcal{S}_j(\mathbf{X})$ are for different groups of users.
Let $S_i(\mathbf{X})$ denote the number of tiles in set $\mathcal{S}_i(\mathbf{X})$.
Let $\mathcal K_{i}(\mathbf X)$ and $K_{i}(\mathbf{X})$ denote the set and the number of users that need to receive the tiles in $\mathcal{S}_i(\mathbf{X})$.
Without
loss of generality, we refer to the transmission of the tiles in $\mathcal{S}_i(\mathbf{X})$ to the users in $\mathcal{K}_{i}(\mathbf{X})$ as multicast, although both unicast ($K_i(\mathbf{X})=1$) and multicast ($K_i(\mathbf{X})>1$) may happen.

\textcolor{black}{We consider a discrete time narrow band TDMA\footnote{\textcolor{black}{\textcolor{black}{Note that TDMA is analytically tractable and has  applications in WiFi systems. In addition, the} multicast transmission scheme and the optimization framework \textcolor{black}{for a TDMA system} can be extended to an OFDMA system.}} system with bandwidth $B$ (in Hz), and study one time frame of duration $T$ (in seconds). Assume block fading, i.e., each channel state (over bandwidth $B$) does not change within the considered time frame.} \textcolor{black}{In addition, we assume that the channel states of $K$ users are random.} Let $H_k\in \mathcal H$ denote the random channel state of user $k$, representing the power of the channel  between user $k$ and \textcolor{black}{the server},  where $\mathcal H$ denotes the finite channel  state space.
Let $\mathbf H\triangleq (H_k)_{k\in\mathcal K}\in \mathcal H^{K}$ denote the random system channel state.

The random system state consists of  the random system viewing direction state $\mathbf X$ and the random system channel state $\mathbf H$, denoted by $(\mathbf X, \mathbf H)\in  \mathcal X^K\times  \mathcal H^K$.\footnote{\textcolor{black}{Note that the analysis and optimization in this letter do not require the independence between $\mathbf{X}$ and $\mathbf{H}$.}} \textcolor{black}{The probability that the random system state $(\mathbf X,\mathbf H)$ takes the value $(\mathbf x,\mathbf h)\in  \mathcal X^K\times  \mathcal H^K$ is denoted as
$\Pr[(\mathbf X,\mathbf H)=(\mathbf x, \mathbf h) ]$,}
\textcolor{black}{where $\mathbf x\triangleq (\mathbf{x}_k)_{k\in\mathcal K}$, $\mathbf{x}_k\in \mathcal X$, $\mathbf h \triangleq (h_k)_{k\in\mathcal K}$, and $h_k\in\mathcal{H}$.}
We assume that \textcolor{black}{the server} is aware of the system state $(\mathbf X,\mathbf H)$.

The time   allocated to transmit the tiles in $\mathcal{S}_i(\mathbf{X})$ is denoted by $t_{i}(\mathbf{X},\mathbf{H})$.
Thus, we have the following transmission time allocation constraints:
\begin{align}
&t_i(\mathbf{X},\mathbf{H})\geq0,~i\in\mathcal{I}(\mathbf{X}),\mathbf{X}\in\mathcal{X}^K,\mathbf{H}\in\mathcal{H}^K,\label{const:tdma-time0}\\
&\sum\nolimits_{i\in\mathcal{I}(\mathbf{X})}t_{i}(\mathbf{X},\mathbf{H})\leq T,~\mathbf{X}\in\mathcal{X}^K,\mathbf{H}\in\mathcal{H}^K.\label{const:tdma-time}
\end{align}
Let $p_{i}(\mathbf{X},\mathbf{H})\geq0$ denote the transmission power of the symbols for the tiles in $\mathcal{S}_i(\mathbf{X})$.
Thus,
the  transmission energy  is:
\begin{align}\label{eqn:tdma-total energy}
E(\mathbf{t}(\mathbf{X},\mathbf{H}),\mathbf{p}(\mathbf{X},\mathbf{H}))=\sum\nolimits_{i\in\mathcal{I}(\mathbf{X})}t_{i}(\mathbf{X},\mathbf{H})p_{i}(\mathbf{X},\mathbf{H}),\nonumber\\
~\mathbf{X}\in\mathcal{X}^K,\mathbf{H}\in\mathcal{H}^K,
\end{align}
where
$\mathbf{t}(\mathbf{X},\mathbf{H})\triangleq\left(t_{i}(\mathbf{X},\mathbf{H})\right)_{i\in\mathcal{I}(\mathbf{X})}$ and $\mathbf{p}(\mathbf{X},\mathbf{H})\triangleq\left(p_{i}(\mathbf{X},\mathbf{H})\right)_{i\in\mathcal{I}(\mathbf{X})}$.
To obtain design insights, we consider capacity achieving code.
To guarantee that all users in $\mathcal{K}_i(\mathbf{X})$ can successfully receive the tiles in $\mathcal{S}_i(\mathbf{X})$, we have the following transmission rate constraint:
\begin{align}
t_{i}(\mathbf{X},\mathbf{H})B \log_2 \left(1 + \frac{p_{i}(\mathbf{X},\mathbf{H}) H_{k}}{n_0}\right)\geq S_{i}(\mathbf{X})DT,\nonumber\\
~k\in\mathcal{K}_i(\mathbf{X}),i\in\mathcal{I}(\mathbf{X}),\mathbf{X}\in\mathcal{X}^K,\mathbf{H}\in\mathcal{H}^K,\label{const:tdma-size}
\end{align}
where $n_0$ is the power of the complex additive  white Gaussian noise at each receiver. Denote $\mathbf{t}\triangleq\left(\mathbf{t}(\mathbf{X},\mathbf{H})\right)_{(\mathbf{X},\mathbf{H})\in\mathcal{X}^K\times\mathcal{H}^K}$ and
$\mathbf{p}\triangleq\left(\mathbf{p}(\mathbf{X},\mathbf{H})\right)_{(\mathbf{X},\mathbf{H})\in\mathcal{X}^K\times\mathcal{H}^K}$.

\section{\textcolor{black}{Average Transmission Energy Minimization}}\label{sec:energy}
\subsection{Problem Formulation}
Given the video quality (i.e., encoding rate of each tile $D\in\{D_1,\ldots,D_L\}$),
we would like to minimize the \textcolor{black}{average transmission energy}  under the transmission time allocation constraints and the transmission rate constraint.
\begin{Prob} [Average  Transmission Energy Minimization]\label{P1}
\begin{align}
 \bar{E}^{\star}\triangleq\min_{\mathbf{t},\mathbf{p}}~&\mathbb{E}[E(\mathbf{t}(\mathbf{X},\mathbf{H}),\mathbf{p}(\mathbf{X},\mathbf{H}))]\nonumber\\
    \mathrm{s.t.} ~~&\eqref{const:tdma-time0},~\eqref{const:tdma-time},~\eqref{const:tdma-size},\nonumber
\end{align}
where the expectation $\mathbb{E}$ is taken over the random system state $(\mathbf{X},\mathbf{H})\in\mathcal{X}^K\times\mathcal{H}^K$, \textcolor{black}{i.e.,
$\mathbb{E}[E(\mathbf{t}(\mathbf{X},\mathbf{H}),\mathbf{p}(\mathbf{X},\mathbf{H}))]=\sum\nolimits_{(\mathbf{x},\mathbf{h})\in\mathcal{X}^K\times\mathcal{H}^K}\Pr[(\mathbf X,\mathbf H)=(\mathbf x, \mathbf h)]E(\mathbf{t}(\mathbf{x},\mathbf{h}),\mathbf{p}(\mathbf{x},\mathbf{h}))$.}
\textcolor{black}{Let $\mathbf{t}^{\star}_{\text{e}}$ and $\mathbf{p}^{\star}_{\text{e}}$ denote an optimal solution.}
\end{Prob}

Note that the objective function of Problem~\ref{P1} and the constraint in $\eqref{const:tdma-size}$ are non-convex. Thus, Problem~\ref{P1} is non-convex.
\textcolor{black}{In general, it is challenging to obtain a globally optimal solution of a non-convex problem.}

\subsection{Optimal Solution}
 In this part, we obtain a globally optimal solution of the non-convex Problem~\ref{P1}.
\textcolor{black}{Note that Problem~\ref{P1} can be \textcolor{black}{decomposed} into subproblems, \textcolor{black}{one for each
 $(\mathbf{X},\mathbf{H})\in\mathcal{X}^K\times\mathcal{H}^K$.}
\begin{align}&\emph{Subproblem~1 (\text{Subproblem of Problem~\ref{P1}):}}\nonumber\\
 &E^{\star}(\mathbf{X},\mathbf{H})\triangleq\min_{\mathbf{t}(\mathbf{X},\mathbf{H}),\mathbf{p}(\mathbf{X},\mathbf{H})}~E(\mathbf{t}(\mathbf{X},\mathbf{H}),\mathbf{p}(\mathbf{X},\mathbf{H}))\nonumber\\
    &\quad\quad\quad\quad\quad\quad\quad\quad\mathrm{s.t.} ~~~\eqref{const:tdma-time0},~\eqref{const:tdma-time},~\eqref{const:tdma-size}.\nonumber
\end{align}}
~Denote $H_{i,\text{min}}(\mathbf{X},\mathbf{H})\triangleq\min\limits_{k\in\mathcal{K}_{i}(\mathbf{X})}H_k(\mathbf{X},\mathbf{H})$.
To solve \textcolor{black}{Subproblem~1}, we first analyze its optimality properties.
\begin{Lem}[Optimality Properties of \textcolor{black}{Subproblem~1}]\label{lem:opt structure}
The optimal solution of \textcolor{black}{Subproblem~1} satisfies:
\begin{align}
p_{\text{e},i}^{\star}(\mathbf{X},\mathbf{H})=\frac{n_0}{H_{i,\text{min}}(\mathbf{X},\mathbf{H})}\left(2^{\frac{S_{i}(\mathbf{X})DT}{Bt^{\star}_{\text{e},i}(\mathbf{X},\mathbf{H})}}-1\right),~i\in\mathcal{I}(\mathbf{X}).\label{eqn:tdma-p1}
\end{align}
\end{Lem}
\textcolor{black}{\begin{proof}[Proof (sketch)]
Suppose an optimal solution of Subproblem~1 is $((t^{\star}_{\text{e},i}(\mathbf{X},\mathbf{H}))_{i\in\mathcal{I}(\mathbf{X})},(p^{\dagger}_{\text{e},i}(\mathbf{X},\mathbf{H}))_{i\in\mathcal{I}(\mathbf{X})})$, and $(p^{\dagger}_{\text{e},i}(\mathbf{X},\mathbf{H}))_{i\in\mathcal{I}(\mathbf{X})}\neq(p^{\star}_{\text{e},i}(\mathbf{X},\mathbf{H}))_{i\in\mathcal{I}(\mathbf{X})}$, \textcolor{black}{where $(p^{\star}_{\text{e},i}(\mathbf{X},\mathbf{H}))_{i\in\mathcal{I}(\mathbf{X})}$} is given by \eqref{eqn:tdma-p1}. Then, there exists $j\in\mathcal{I}(\mathbf{X})$ such that $p^{\dagger}_{\text{e},j}(\mathbf{X},\mathbf{H})<p^{\star}_{\text{e},j}(\mathbf{X},\mathbf{H})$.
\textcolor{black}{By} \eqref{eqn:tdma-p1}, we can show that $p^{\dagger}_{\text{e},j}(\mathbf{X},\mathbf{H})$ does not satisfy  \eqref{const:tdma-size}, which \textcolor{black}{contradicts with the assumption}.
Therefore, \textcolor{black}{by contradiction, we can prove} Lemma~\ref{lem:opt structure}.
\end{proof}}

\textcolor{black}{Note that $\frac{S_{i}(\mathbf{X})DT}{t^{\star}_{\text{e},i}(\mathbf{X},\mathbf{H})}$ in \eqref{eqn:tdma-p1} represents the transmission rate (in bit/s) for the tiles in $\mathcal{S}_i(\mathbf{X})$ for given $(\mathbf{X},\mathbf{H})$.}
By Lemma~\ref{lem:opt structure}, we can eliminate $\mathbf{p}$ and transform \textcolor{black}{Subproblem~1} into an equivalent problem:

\begin{Prob} [Equivalent Problem of \textcolor{black}{Subproblem~1}]\label{ESP1}
\begin{align}
 \min_{\textcolor{black}{\mathbf{t}(\mathbf{X},\mathbf{H})}}~&\textcolor{black}{\sum\nolimits_{i\in\mathcal{I}(\mathbf{X})}\frac{n_0t_{i}(\mathbf{X},\mathbf{H})}{H_{i,\text{min}}(\mathbf{X},\mathbf{H})}\left(2^{\frac{S_{i}(\mathbf{X})DT}{Bt_{i}(\mathbf{X},\mathbf{H})}}-1\right)}\nonumber\\
    \mathrm{s.t.}~~&\eqref{const:tdma-time0},~\eqref{const:tdma-time}.\nonumber
\end{align}
\end{Prob}

We can easily verify that Problem~\ref{ESP1} is convex and the Slater's condition is satisfied, implying that strong duality holds. Thus, Problem~\ref{ESP1} can be solved using KKT conditions \textcolor{black}{as in \cite[pp.~243-246]{cvx}}. Based on the optimal solution of Problem~\ref{ESP1} and Lemma~\ref{lem:opt structure}, we have the following result.
\begin{Lem}[Optimal Solution  of Problem~\ref{P1}]\label{lem:opt solution}
\begin{align}
t_{\text{e},i}^{\star}(\mathbf{X},\mathbf{H})=&\frac{S_{i}(\mathbf{X})DT\ln{2}}{B\left(W\left(\frac{\lambda^{\star}(\mathbf{X},\mathbf{H})H_{i,\text{min}}(\mathbf{X},\mathbf{H})}{n_0e}-\frac{1}{e}\right)+1\right)},\nonumber\\
&~i\in\mathcal{I}(\mathbf{X}),\mathbf{X}\in\mathcal{X}^K,\mathbf{H}\in\mathcal{H}^K,\label{eqn:tdma-opt solution1}\\
p_{\text{e},i}^{\star}(\mathbf{X},\mathbf{H})=&\frac{n_0\left(e^{W\left(\frac{\lambda^{\star}(\mathbf{X},\mathbf{H})H_{i,\text{min}}(\mathbf{X},\mathbf{H})}{n_0e}-\frac{1}{e}\right)+1}-1\right)}{H_{i,\min}(\mathbf{X},\mathbf{H})},\nonumber\\
&~i\in\mathcal{I}(\mathbf{X}),\mathbf{X}\in\mathcal{X}^K,\mathbf{H}\in\mathcal{H}^K,\label{eqn:tdma-opt solution11}
\end{align}
where $W(\cdot)$ denotes the Lambert function, $e$ is a mathematical constant that is the base of the natural logarithm, and the Lagrangian multiplier $\lambda^{\star}(\mathbf{X},\mathbf{H})$ satisfies
\begin{align}\label{eqn:tdma-lambda1}
\sum\nolimits_{i\in\mathcal{I}(\mathbf{X})}\frac{S_{i}(\mathbf{X})D\ln{2}}{B\left(W\left(\frac{\lambda^{\star}(\mathbf{X},\mathbf{H})H_{i,\text{min}}(\mathbf{X},\mathbf{H})}{n_0e}-\frac{1}{e}\right)+1\right)}=1,\nonumber\\
~\mathbf{X}\in\mathcal{X}^K,\mathbf{H}\in\mathcal{H}^K.
\end{align}
\end{Lem}

Note that $\lambda^{\star}(\mathbf{X},\mathbf{H})$ in \eqref{eqn:tdma-lambda1} can be easily obtained using the bisection method. Thus, we can obtain a globally optimal solution of Problem~\ref{P1} efficiently.
\textcolor{black}{
By Lemma~\ref{lem:opt structure} and Lemma~\ref{lem:opt solution}, we have the following result.}
\begin{Lem}[Minimum Transmission Energy]\label{lem:bound}
\textcolor{black}{If $H_{1}=\cdots=H_{K}\triangleq \widetilde{H}$, then
\begin{align}
E^{\star}(\mathbf{X},\mathbf{H})=\frac{n_0T}{\widetilde{H}}\left(2^{\frac{D\sum\nolimits_{i\in\mathcal{I}(\mathbf{X})}S_i(\mathbf{X})}{B}}-1\right).\label{pro:energy}
\end{align}
Furthermore,} for all $(\mathbf{X},\mathbf{H})\in\mathcal{X}^K\times\mathcal{H}^K$,
\begin{align}\label{eqn:bound}
&\frac{n_0T}{\max \mathcal{H}}\left(2^{\frac{D\sum\nolimits_{i\in\mathcal{I}(\mathbf{X})}S_i(\mathbf{X})}{B}}-1\right)\leq E^{\star}(\mathbf{X},\mathbf{H})\nonumber\\
&\leq\frac{n_0T}{\min\mathcal{H}}\left(2^{\frac{D\sum\nolimits_{i\in\mathcal{I}(\mathbf{X})}S_i(\mathbf{X})}{B}}-1\right).
\end{align}
\end{Lem}
\textcolor{black}{\begin{proof}[Proof (sketch)]
When $H_{1}=\cdots=H_{K}=\widetilde{H}$,
by Lemma~\ref{lem:opt solution} and \eqref{eqn:tdma-total energy}, we \textcolor{black}{can prove \eqref{pro:energy}.}
For given $\mathbf{X}$, we can show that \textcolor{black}{any} optimal solution of Subproblem~1 with arbitrary $\mathbf{H}$ \textcolor{black}{is a feasible solution} of Subproblem~1 with $H_{1}=\cdots=H_{K}=\max\mathcal{H}$, \textcolor{black}{and any optimal solution of Subproblem~1 with $H_{1}=\cdots=H_{K}=\min\mathcal{H}$ is a feasible solution of Subproblem~1 with arbitrary $\mathbf{H}$.} Thus, by \eqref{pro:energy}, we \textcolor{black}{can prove \eqref{eqn:bound}.}
\end{proof}}

\textcolor{black}{Lemma~\ref{lem:bound} indicates that approximately, the minimum transmission energy $E^{\star}(\mathbf{X},\mathbf{H})$ increases  \textcolor{black}{exponentially} with  \textcolor{black}{the total number of tiles that need to be transmitted, i.e., $\sum\nolimits_{i\in\mathcal{I}(\mathbf{X})}S_i(\mathbf{X})$, and is inversely proportional to the channel \textcolor{black}{powers}, i.e., $H_k,k\in\mathcal{K}$.}} Note that $\sum\nolimits_{i\in\mathcal{I}(\mathbf{X})}S_i(\mathbf{X})$ reflects the concentration of the viewing  directions of all users.
A smaller value of $\sum\nolimits_{i\in\mathcal{I}(\mathbf{X})}S_i(\mathbf{X})$ means closer viewing directions of all users.

\section{Received Video Quality Maximization}\label{sec:quality}
\subsection{Problem Formulation}
Let $E_{\text{limit}}$ denote the  transmission  energy budget of the system.  Consider the maximum transmission energy  constraint:
\begin{align}\label{const:tdma-energy constraint}
\sum\nolimits_{i\in\mathcal{I}(\mathbf{X})}t_{i}(\mathbf{X},\mathbf{H})p_{i}(\mathbf{X},\mathbf{H})\leq E_{\text{limit}},~\mathbf{X}\in\mathcal{X}^K,\mathbf{H}\in\mathcal{H}^K.
\end{align}
To guarantee user experience, the encoding rate should not change as frequently as the viewing  directions and channel states, and should remain constant within a certain time \textcolor{black}{duration}.
Given the \textcolor{black}{transmission energy budget} $E_{\text{limit}}$, we would like to maximize the received video quality (i.e., encoding rate of each tile $D$) under the maximum transmission energy  constraint.
\begin{Prob} [Received Video Quality Maximization]\label{P2}
\begin{align}
D_{\text{q}}^{\star}&\triangleq\max_{D,\mathbf{t},\mathbf{p}}D\nonumber\\
    \mathrm{s.t.} ~~\quad
    &\eqref{const:tdma-time0},~\eqref{const:tdma-time},~\eqref{const:tdma-size},~\eqref{const:tdma-energy constraint}.\nonumber
\end{align}
Let $D_{\text{q}}^{\star}$, $\mathbf{t}_{\text{q}}^{\star}$ and $\mathbf{p}_{\text{q}}^{\star}$ denote an optimal solution.
\end{Prob}

Note that the actual achievable video quality is $\max\limits_{l\in\{1,\ldots,L\}}\{D_l|D_l\leq D^{\star}_{\text{q}}\}$.
The constraints in $\eqref{const:tdma-size}$ and \eqref{const:tdma-energy constraint} are non-convex. Thus, Problem~\ref{P2} is non-convex.

\subsection{Optimal Solution}
In this part, we obtain a globally optimal solution of the non-convex Problem~\ref{P2}. First, we transform it into an equivalent convex problem.
Let $\mathbf{H}_{\text{min}}\triangleq(H_k)_{k\in\mathcal{K}}$ with $H_k=\min\mathcal{H}$.
\begin{Prob} [Equivalent Problem of Problem~\ref{P2}]\label{EP2}
\begin{align}
&\min_{\mathbf{X}\in\mathcal{X}^K}D_{\text{q}}^{\star}(\mathbf{X},\mathbf{H}_{\min})\nonumber
\end{align}
where $D_{\text{q}}^{\star}(\mathbf{X},\mathbf{H}_{\min})$ is given by the following subproblem.
\end{Prob}
\begin{Prob} [Subproblem of Problem~\ref{EP2}]\label{SEP2}
For all $\mathbf X\in \mathcal X^K$,
\begin{align}
&D^{\star}_{\text{q}}(\mathbf{X},\mathbf{H}_{\min})\triangleq\max_{D,\mathbf{t}(\mathbf{X},\mathbf{H}_{\min})}D\nonumber\\
    &\mathrm{s.t.} \quad  \eqref{const:tdma-time0},~\eqref{const:tdma-time},\nonumber\\
    & \sum\nolimits_{i\in\mathcal{I}(\mathbf{X})}\frac{n_0t_{i}(\mathbf{X},\mathbf{H}_{\min})}{\min\mathcal{H}}\left(2^{\frac{S_{i}(\mathbf{X})DT}{Bt_{i}(\mathbf{X},\mathbf{H}_{\min})}}-1\right)\leq E_{\text{limit}}.\label{const:tdma-energy const}
\end{align}
\end{Prob}
By carefully exploring structural properties of Problem~\ref{P2}, we have the following result.
\begin{Lem}\label{lem:equivalent}
\textcolor{black}{The optimal values of  Problem~\ref{P2} and Problem~\ref{EP2} are equivalent, i.e., $D^{\star}_{\text{q}}=\min\limits_{\mathbf{X}\in\mathcal{X}^K}D^{\star}_{\text{q}}(\mathbf{X},\mathbf{H}_{\min})$.}
Furthermore,
\begin{align}
p_{\text{q},i}^{\star}(\mathbf{X},\mathbf{H}_{\min})=\frac{n_0}{\min\mathcal{H}}\left(2^{\frac{S_{i}(\mathbf{X})D_{\text{q}}^{\star}(\mathbf{X},\mathbf{H}_{\min})T}{Bt^{\star}_{\text{q},i}(\mathbf{X},\mathbf{H}_{\min})}}-1\right), \nonumber\\
 i\in\mathcal{I}(\mathbf{X}),\mathbf{X}\in\mathcal{X}^K.\label{eqn:tdma-p2}
\end{align}
\end{Lem}

\textcolor{black}{Note that Lemma~\ref{lem:equivalent} can be proved in a similar way to  Lemma~\ref{lem:opt structure}.}
We can easily verify that Problem~\ref{SEP2} is convex and the Slater's condition is satisfied, implying that strong duality holds. Thus, Problem~\ref{SEP2} can be solved using KKT conditions \textcolor{black}{as in \cite[pp.~243-246]{cvx}}. Based on the optimal \textcolor{black}{solution} of Problem~\ref{SEP2} and Lemma~\ref{lem:equivalent}, we have the following result.

\begin{Lem}[Optimal Solution of Problem~\ref{P2}]\label{lem:quality opt structure}
\begin{align}
&D_{\text{q}}^{\star}=\frac{B\ln(\frac{E_{\text{limit}}\min\mathcal{H}}{n_0T}+1)}{\ln{2}\max\limits_{\mathbf{X}\in\mathcal{X}^K}\sum\nolimits_{i\in\mathcal{I}(\mathbf{X})}S_{i}(\mathbf{X})},\label{eqn:tdma-opt D}
\end{align}
and $t_{\text{q},i}^{\star}(\mathbf{X},\mathbf{H})$ and $p_{\text{q},i}^{\star}(\mathbf{X},\mathbf{H})$ are given by $t_{\text{e},i}^{\star}(\mathbf{X},\mathbf{H})$ in \eqref{eqn:tdma-opt solution1} and $p_{\text{e},i}^{\star}(\mathbf{X},\mathbf{H})$ in \eqref{eqn:tdma-opt solution11}, respectively, with $D=D_{\text{q}}^{\star}$.
\end{Lem}

\textcolor{black}{Lemma~\ref{lem:quality opt structure} indicates that  $D_{\text{q}}^{\star}$ is affected by \textcolor{black}{the smallest channel power $\min\mathcal{H}$ among all channel powers}  instead of $H_i,i\in\mathcal{I}(\mathbf{X})$, and is inversely proportional to $\max\limits_{\mathbf{X}\in\mathcal{X}^K}\sum\nolimits_{i\in\mathcal{I}(\mathbf{X})}S_i(\mathbf{X})$ which represents the maximum number of tiles that need to be transmitted for all viewing directions.}

\section{Simulation}\label{sec:simulation}

\begin{figure}[t]
\vspace*{-0.65cm}
\begin{center}
 \subfigure[\small{Average transmission energy versus $\gamma$. $K=3$, $D=30561$ bit/s, $d=1\times10^{-6}$.}]
 {\resizebox{4.11cm}{!}{\includegraphics{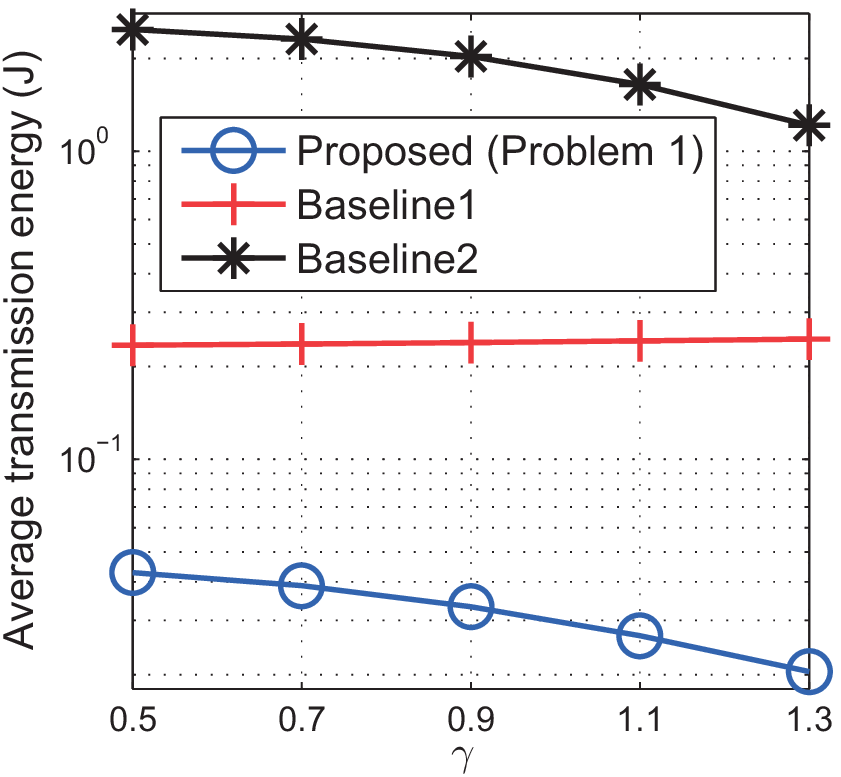}}}
  \subfigure[\small{Encoding rate of each tile versus $d$. $K=4$, $\gamma=0.8$, $E_{\text{limit}}=0.1$ J.}]
 {\resizebox{4.17cm}{!}{\includegraphics{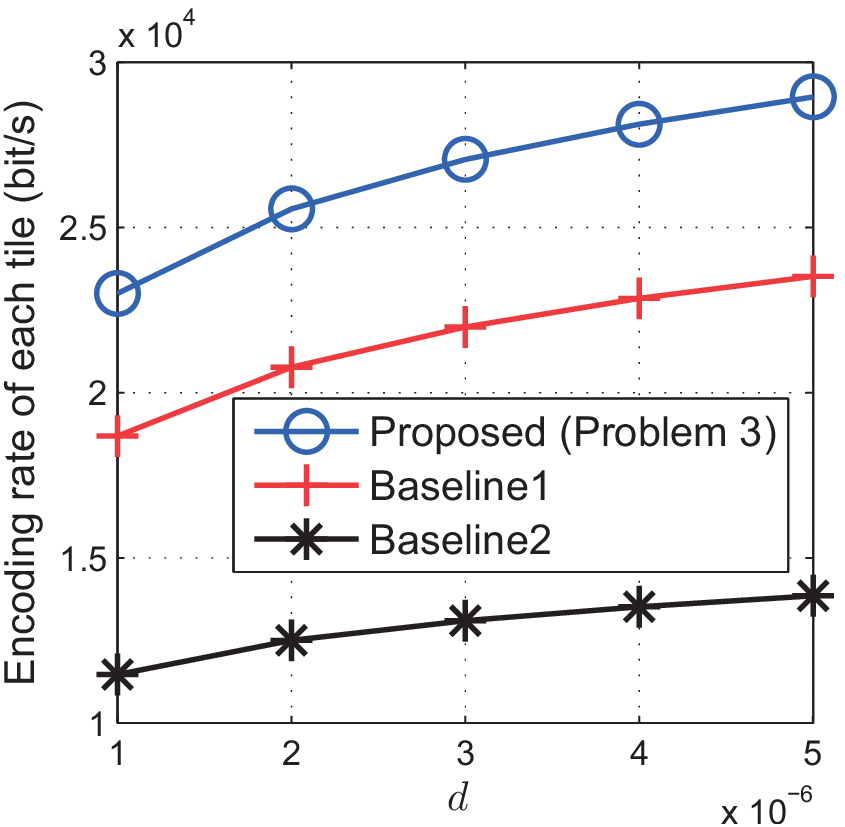}}}
 \vspace*{-0.42cm}
 \end{center}
   \caption{\small{Performance comparison.}}
   \label{fig:energy}
\vspace*{-0.55cm}
\end{figure}

\textcolor{black}{In this section, we compare the proposed solutions in Section~\ref{sec:energy} and Section~\ref{sec:quality} with two baselines using numerical results.
\textcolor{black}{Baseline 1 considers serving each user separately using unicast in an optimal way, similar to the proposed optimal solutions. Baseline~2 considers  multicast as in this letter but with equal time allocation for each transmitted tile and optimal power allocation based on the equal time allocation.}
In the simulation, we use Kvazaar as the 360 VR video encoder and video sequence $Boxing$ as the video source \cite{boxing}. We set $F_h=F_v=100^{\circ}$, \textcolor{black}{$N_h\times N_v=36\times2$}, and \textcolor{black}{$V_h\times V_v=30\times15$}. To avoid view switch delay, we transmit extra $15^{\circ}$ in the four directions of each requested FoV.
Set $B =10$ MHz, $n_0 = 10^{-9}$ W, $T=0.1$ s,
$\mathcal{H}=\{0.5d,1.5d\}$, \textcolor{black}{$\Pr[H_k=0.5d]=0.5$, and  $\Pr[H_k=1.5d]=0.5$} for all $k\in\mathcal{K}$, where $d$ reflects the path loss.
For ease of exposition, we consider Zipf distribution for the $N_h\times N_v$ viewing directions. In particular, suppose viewing direction $(n_h,n_v)$ is of rank\footnote{\textcolor{black}{Note that $(n_h,n_v)$ \textcolor{black}{can be an
arbitrary rank}, and the ranks of the viewing directions do not influence the trends of the curves in Fig.~\ref{fig:energy} (a), but the Zipf exponent $\gamma$ does. In addition, the ranks of the viewing directions do not influence the curves in Fig.~\ref{fig:energy}~(b), as the solution of Problem~\ref{P2} depends only on $\mathcal{X}^K\times\mathcal{H}^K$.}} $(n_h-1)N_v+n_v$ and $\Pr[\mathbf{X}_k=(n_h,n_v)]=\frac{((n_h-1)N_v+n_v)^{-\gamma}}{\sum\nolimits_{i=1,\ldots,N_hN_v}i^{-\gamma}}$, where $\gamma$ is the Zipf exponent. Note that a smaller $\gamma$ indicates a longer tail.}

\textcolor{black}{Fig.~\ref{fig:energy} (a) illustrates the average transmission energy versus the Zipf exponent $\gamma$.
\textcolor{black}{\textcolor{black}{We can see that the} average transmission energy of the two multicast schemes decreases with $\gamma$,} \textcolor{black}{as multicast opportunities increases with $\gamma$.}
Fig.~\ref{fig:energy} (b) illustrates the received video quality (i.e., encoding rate of each tile) versus the path loss $d$.  \textcolor{black}{\textcolor{black}{We can observe that the} received video quality of each scheme increases with the channel powers.}
From Fig.~\ref{fig:energy}, we can \textcolor{black}{also} observe that the proposed optimal solutions outperform the two baselines. Specifically,  the gains of the proposed optimal solutions over Baseline~1 arise from the fact that the proposed solutions utilize multicast.
The gains of the proposed optimal solutions over Baseline~2 are attributed by the fact that the proposed  solutions carefully allocate both transmission time and transmission power.}

\section{Conclusion}
In this letter, we \textcolor{black}{studied} optimal multicast of tiled 360 VR video
using TDMA.
We \textcolor{black}{considered} random
viewing directions and random channel conditions\textcolor{black}{. We} \textcolor{black}{formulated} two non-convex
optimization problems, i.e., the minimization of the average
transmission energy for given video
quality, and the maximization of the received video quality
 for given
\textcolor{black}{transmission energy budget}. We  \textcolor{black}{obtained}  globally optimal closed-form solutions of the two challenging non-convex problems and \textcolor{black}{revealed} important design insights for tiled 360 VR multicast. Finally, numerical results
\textcolor{black}{demonstrated} the advantage of the proposed optimal solutions.

\bibliographystyle{IEEEtran}

\begin{thebibliography}{1}
\providecommand{\url}[1]{#1}
\csname url@samestyle\endcsname
\providecommand{\newblock}{\relax}
\providecommand{\bibinfo}[2]{#2}
\providecommand{\BIBentrySTDinterwordspacing}{\spaceskip=0pt\relax}
\providecommand{\BIBentryALTinterwordstretchfactor}{4}
\providecommand{\BIBentryALTinterwordspacing}{\spaceskip=\fontdimen2\font plus
\BIBentryALTinterwordstretchfactor\fontdimen3\font minus
  \fontdimen4\font\relax}
\providecommand{\BIBforeignlanguage}[2]{{%
\expandafter\ifx\csname l@#1\endcsname\relax
\typeout{** WARNING: IEEEtran.bst: No hyphenation pattern has been}%
\typeout{** loaded for the language `#1'. Using the pattern for}%
\typeout{** the default language instead.}%
\else
\language=\csname l@#1\endcsname
\fi
#2}}
\providecommand{\BIBdecl}{\relax}
\BIBdecl

\bibitem{111}
``\protect{Augmented/Virtual Reality} revenue forecast revised to hit 120
  billion by 2020,'' \url{https://goo.gl/nw9mtP}, Jan. 2016.

\bibitem{unicast-liu}
Z.~Liu, S.~Ishihara, Y.~Cui, Y.~Ji, and Y.~Tanaka, ``Jet: Joint source and
  channel coding for error resilient virtual reality video wireless
  transmission,'' \emph{Signal Processing}, vol. 147, pp. 154--162, Jun. 2018.

\bibitem{unicast-one}
\BIBentryALTinterwordspacing
S.~Xie, Q.~Shen, Y.~Xu, Q.~Qian, S.~Zhang, Z.~Ma, and W.~Zhang, ``Viewport
  adaptation-based immersive video streaming: Perceptual modeling and
  applications,'' \emph{CoRR}, vol. abs/1802.06057, 2018. [Online]. Available:
  \url{http://arxiv.org/abs/1802.06057}
\BIBentrySTDinterwordspacing

\bibitem{multicast-nojoint}
H.~Ahmadi, O.~Eltobgy, and M.~Hefeeda, ``Adaptive multicast streaming of
  virtual reality content to mobile users,'' in \emph{Proc. Proceedings of the
  on Thematic Workshops of ACM Multimedia}, Oct. 2017, pp. 170--178.

\bibitem{multicast-olcoding}
Y.~Bao, T.~Zhang, A.~Pande, H.~Wu, and X.~Liu, ``Motion-prediction-based
  multicast for 360-degree video transmissions,'' in \emph{Proc. IEEE
  International Conference on Sensing, Communication, and Networking (SECON)},
  Jun. 2017, pp. 1--9.

\bibitem{cvx}
S.~Boyd and L.~Vandenberghe, \emph{Convex optimization}.\hskip 1em plus 0.5em
  minus 0.4em\relax Cambridge university press, 2004.

\bibitem{boxing}
\url{http://www.utovr.com/video/101611104210.html}.

\end{thebibliography}

\end{document}